\documentclass[aps,prl,twocolumn,groupedaddress,amsmath,showpacs,superscriptaddress]{revtex4-1}
\usepackage{amsmath}
\usepackage{graphicx}
\usepackage{amsfonts}
\usepackage{amssymb}
\usepackage{hyperref}
\usepackage{color}
\usepackage{mathrsfs}
\usepackage{isomath}
\usepackage{amsmath}
\usepackage{amsthm}

\usepackage{txfonts}

\newcommand{\assem}[3]{#1_{#2}^{#3}}
\newcommand{\rB}{{B}}
\newcommand{\rA}{{A}}
\newcommand{\Tr}{\mbox{Tr}}
\newcommand{\Exp}[1]{\langle #1\rangle}
\newcommand{\Expt}[2]{\langle #1 \otimes #2\rangle}

\begin{document}
\title{Hierarchy of Steering Criteria Based on Moments for All Bipartite Quantum Systems}

\author{Ioannis Kogias}
\email{john$_$k$_$423@yahoo.gr}
\affiliation{$\mbox{School of Mathematical Sciences, The University of Nottingham,
University Park, Nottingham NG7 2RD, United Kingdom}$}
\affiliation{$\mbox{ICFO - The Institute of Photonic Sciences, Av. Carl Friedrich Gauss, 3 08860 Castelldefels (Barcelona), Spain}$}

\author{Paul Skrzypczyk}
\email{paul.skrzypczyk@bristol.ac.uk}
\affiliation{$\mbox{H. H. Wills Physics Laboratory, University of Bristol, Tyndall Avenue, Bristol, BS8 1TL, United Kingdom}$}
\affiliation{$\mbox{ICFO - The Institute of Photonic Sciences, Av. Carl Friedrich Gauss, 3 08860 Castelldefels (Barcelona), Spain}$}

\author{Daniel Cavalcanti}
\email{daniel.cavalcanti@icfo.es}
\affiliation{$\mbox{ICFO - The Institute of Photonic Sciences, Av. Carl Friedrich Gauss, 3 08860 Castelldefels (Barcelona), Spain}$}

\author{Antonio Ac{\'i}n}
\email{antonio.acin@icfo.es}
\affiliation{$\mbox{ICFO - The Institute of Photonic Sciences, Av. Carl Friedrich Gauss, 3 08860 Castelldefels (Barcelona), Spain}$}

\author{Gerardo Adesso}
\email{Gerardo.Adesso@nottingham.ac.uk}
\affiliation{$\mbox{School of Mathematical Sciences, The University of Nottingham,
University Park, Nottingham NG7 2RD, United Kingdom}$}


\begin{abstract}

Einstein-Podolsky-Rosen steering is a manifestation of quantum correlations exhibited by quantum systems, that allows for entanglement certification when one of the subsystems is not characterized. Detecting steerability of quantum states is essential to assess their suitability for quantum information protocols with partially trusted devices. We provide a hierarchy of sufficient conditions for the steerability of bipartite quantum states of any dimension, including continuous variable states. Previously known steering criteria are recovered as special cases of our approach. The proposed method allows us to derive optimal steering witnesses for arbitrary families of quantum states, and provides a systematic framework to analytically derive non-linear steering criteria. We discuss relevant examples and, in particular, provide an optimal steering witness for a lossy single-photon Bell state; the witness can be implemented just by linear optics and homodyne detection, and detects steering with a higher loss tolerance than any other known method.  Our approach is readily applicable to multipartite steering detection and to the characterization of joint measurability.

\end{abstract}

\date{\today}
\maketitle

\textit{Introduction ---} Entanglement certification \cite{guhnetoth} is a major topic within quantum information science, as entanglement is behind many applications \cite{chuaniels}. In this task, the knowledge of the system's dimension and the assumption that the measuring devices  are trusted, i.e.~they operate as prescribed, are usually taken for granted;  yet a breakdown of any of these could undermine the entanglement certification \cite{entcertif}. Motivated by quantum cryptography, where minimal assumptions are desired, it was  realized that nonlocal entangled states, i.e.~states that violate a Bell inequality \cite{bell64,bell76,nonloc}, allow for entanglement certification with systems and devices being completely uncharacterized \cite{crypto}. Fully-device-independent entanglement certification, however,  requires the observation of Bell violations free of the detection loophole, which is experimentally very demanding \cite{hensen}.

Einstein-Podolsky-Rosen  steering \cite{schr,schr2,wiseman} is a type of quantum correlation that is intermediate between entanglement and nonlocality \cite{saunders,quintinoinequiv}. Witnessing  steering in a bipartite system implies entanglement certification without any assumption on one of the parties, i.e.~in a one-sided device-independent manner. Being less stringent than nonlocality, steering is more robust against experimental noise \cite{saunders,handchen,smith,bennet,Fuwa}, and its loophole-free detection is feasible with present-day technology \cite{wittmann}. Steering also enjoys a plethora of applications, for which plain entanglement is not enough while the harder-to-get nonlocality is not required. These applications range from one-sided device-independent quantum key distribution \cite{branciard}, advantages in subchannel discrimination \citep{PianiSt}, secure quantum teleportation \cite{securetelep}, and  connections to joint measurability of generalized measurements \cite{uola,quintinojoint,losstolerant,uola2015one}.

Compared to well-studied entanglement and nonlocality, relatively little progress has been achieved about steering detection. A handful of criteria exist \cite{wisepra,cavalcanti,walborn11,walborn13,pramanik1,pramanik2,CHSHanalogue,JonesWise}, which are however tailored to specific measurement scenarios.
Only very recently some constructive steering criteria were introduced, which give an experimenter the freedom to choose the measurements involved, and allow for an improvement of the detection by performing additional measurements until a violation is observed \cite{pusey,PianiSt,Skrzypczyk,steeringmaps}. These criteria are based on the useful methods of semidefinite programming \cite{SDP}, and the downside in this case is that, so far, they could only be applied to discrete variable (DV) systems with not too high dimension, due to computational limitations. It is then clear that there exists still a gap that needs to be filled about steering detection, regarding higher dimensional DV systems and general continuous variable (CV) systems.

In this Letter, we propose a hierarchy of  steering criteria that is directly applicable to bipartite quantum systems of any dimension, including the case of infinite-dimensional systems. Our method avoids the dimension problem by utilizing moments of observables instead of dealing with conditional states, at variance with previous DV proposals. A systematic framework is provided for deriving non-linear steering inequalities in an analytical manner. To the best of our knowledge, our proposed method is the first instance of a hierarchical family of criteria for quantum steering that is valid for any dimension, and shares some similarity in spirit and structure with the hierarchy of moments by Shchukin and Vogel \cite{SV} for CV entanglement detection, and with the Navascu\'es-Pironio-Ac\'in  hierarchy \cite{NPA} for the characterization of nonlocal quantum correlations.  We show that our approach provides optimal moment-based linear steering witnesses for any chosen states and measurements on both parties, including CV ones. Furthermore, various previously proposed steering criteria are retrieved as special cases of our unifying approach, while new non-linear criteria are derived. Finally,  we consider several examples of both DV and CV states, and show that our technique allows to beat the current state-of-the-art in steering detection of a lossy single-photon entangled state with quadrature measurements \cite{JonesWise}.

\textit{Steering detection ---}
We consider the entanglement certification task in which two distant parties, Alice and Bob, each holding one half of a quantum state $\rho_{AB}$ of a bipartite system (described by a Hilbert space ${\cal H}_A \otimes {\cal H}_B$, where ${\cal H}_A ,{\cal H}_B $ denote the Hilbert spaces of Alice and Bob respectively), want to verify that they share entanglement. Additionally to this, we impose the constraint that Alice's system is unknown (i.e., unknown ${\cal H}_A$), and her measurement devices cannot be trusted. This implies that the measurement outcomes  Alice announces cannot be assumed to originate from a particular observable on some quantum state of known dimension. The usual entanglement criteria in this case are inapplicable and we need to consider steering criteria to identify any nonseparability between the untrusted Alice and the trusted Bob~\cite{wiseman}.

In this scenario, Alice performs one out of $n$ unknown measurements (often called `inputs') on her half of $\rho_{AB}$, labelled by $x=1,\ldots ,n$, and with probability $p(a|x)$ gets some outcome $a$.
In principle, Alice's measurements are arbitrary, but one can restrict the analysis to projective measurements without losing generality, because the ancilla needed for a non-projective measurement can always be moved to the definition of the local state on Alice's side. Alice announces the corresponding pair $(a,x)$ to Bob, who then tomographically reconstructs his conditional local (unnormalized) state $\assem{\sigma}{a|x}{\rB}$ which is of arbitrary, but known, dimension. Bob's states are defined so that $\Tr(\assem{\sigma}{a|x}{\rB})=p(a|x)$. For all possible pairs $(a,x)$, Bob thus obtains  the set  $\{\assem{\sigma}{a|x}{\rB}\}$, called an `assemblage' \cite{pusey}. From the assemblage alone,  they should judge whether entanglement was present between their shared systems. We refer to this procedure as a \textit{steering test}.

More precisely, based on the observed assemblage, they must determine whether there exists a separable model, i.e., a separable state $\bar{\rho}_{AB} = \sum_\lambda q_\lambda \, \rho^A_\lambda \otimes \rho^B_\lambda$ on ${\cal H}_A^{\star}\otimes {\cal H}_B$, and  measurements $\{M_{a|x}\}_x $ for Alice, that reproduces Bob's assemblage if we allowed for arbitrary Hilbert spaces ${\cal H}_A^{\star}$ on Alice. If such a model does not exist, then the shared state must be entangled.
A steering test using a separable state $\bar{\rho}_{AB}$, and measurements $\{M_{a|x}\}_x $ associated to each input, necessarily leads to the following form for Bob's conditional (unnormalized) states,
\begin{equation}\label{separable}
\assem{\bar{\sigma}}{a|x}{\rB}  = {\Tr}_A [ \left( M_{a|x} \otimes \openone_B \right) \bar{\rho}_{AB} ]  = \sum_\lambda q_\lambda \, p(a|x,\lambda)\, \rho^B_\lambda, \,\, \forall a,x,
\end{equation}
where $p(a|x,\lambda) = \Tr [ M_{a|x}\,\rho^A_\lambda ]$ and $p(a|x) = \Tr [ \assem{\bar{\sigma}}{a|x}{\rB} ]$.
Assemblages of the form \eqref{separable} are called \textit{unsteerable} \cite{wiseman}. One can also prove that, given any unsteerable assemblage, there always exist a separable state and projective  measurements 
 for Alice that reproduce it.
Furthermore Alice's measurements can be assumed to be described by  mutually commuting observables \cite{appendix}. Intuitively, this follows from the fact that a separable model is `classical' on Alice's side. Therefore, unsteerability is equivalent to the existence of such a separable model.

Our approach is based upon the fact that Bob's conditional states, $\assem{\sigma}{a|x}{\rB}$, on which the steering test is based, are in general hard to obtain experimentally when the set of outcomes is large, or even continuous, as Bob would need to do tomography for every pair $(a,x)$. To circumvent this problem we instead consider the more accessible correlations
\begin{equation}\label{moments}
\langle A_x^\varsigma \otimes B_y^\tau \rangle = \sum\limits_{a,b} a^\varsigma\, b^\tau \, P(a,b|x,B_y) = \sum_a a^\varsigma \mbox{Tr}\left[\sigma_{a|x}B_y^\tau\right],
\end{equation}
between the unknown observables $A_x = \sum_a a M_{a|x}$ (with $x=1,\ldots,n$) measured by Alice,  and some known observables $B_y$ on ${\cal H}_B$ (with $y=1,\ldots\,m$) measured by Bob, with outcomes (eigenvalues) $b$. In Eq.~(\ref{moments}),  $\varsigma,\tau\geq 0$ are integer powers, and $P(a,b|x,B_y)$ is the observed joint probability distribution. In what follows we will show how to derive tests for steering, based solely upon the observed correlations $\{\langle A_x^\varsigma \otimes B_y^\tau\rangle\}$. 

\textit{Moment matrices ---} The main tool we will use is a \textit{moment matrix}, defined as a  $k \times k$ matrix ${\bf{\Gamma}}$ with elements
\begin{equation}\label{momentmatrix}
{\Gamma}_{ij} = \langle  S_i^\dagger S_{j} \rangle\,,
\end{equation}
where $i, j = 1,\ldots, k$, and each operator $S_i$ is some (as-yet unspecified) product of operators for Alice and Bob. As a simple example, if Bob's system is a qubit, one could choose the set ${\cal S} = \{ \openone \otimes \openone, A_1 \otimes X, A_2 \otimes Y, A_3 \otimes Z \}$ where Bob's observables $X,Y,Z$ denote the three Pauli operators.

We first remark that such a moment matrix, when constructed from
physical observables on quantum states, is always positive semidefinite, i.e.~${\bf{\Gamma}} \geq 0$. This follows immediately, since for any vector $\mathbf{v}$, with elements $v_i$, $\sum_{ij} v^*_i \langle  S_i^\dagger S_{j} \rangle v_j = \left\langle \left(\sum_i v^*_i S_i^\dagger\right)\left(\sum_j S_j v_j\right)\right\rangle \geq 0$. The second crucial property is that if the underlying operators satisfy any algebraic properties, then the moment matrix inherits additional structure in the form of linear constraints. For example, if two (hermitian) operators commute, $[S_i, S_j] = 0$, then the corresponding elements of the moment matrix are necessarily equal, $\Gamma_{ij} = \langle S_i^\dagger S_j\rangle =  \langle S_j^\dagger S_i\rangle = \Gamma_{ji}$. As a second example, if $S_i^\dagger S_j = i S_k$ and $S_1 = \openone$, then $\Gamma_{ij} = \langle S_i^\dagger S_j \rangle = i \langle \openone^\dagger S_k\rangle = i\Gamma_{1k}$. In the next section we show that these properties allow us to construct a steering test based upon moment matrices.


\textit{{Detection method based on the moment matrix}} ---
Consider a steering test defined by a set of observed correlations \eqref{moments} and take any set of operators ${\cal S}$ involving some unknown operators on Alice's untrusted side and known operators on Bob's trusted side. Now consider the unknown moment matrix ${\bf{\Gamma}}$ associate to ${\cal S}$ defined as in Eq.~\eqref{momentmatrix}. Some of its matrix elements however are known as they correspond directly to observable data in the steering scenario: these include moments of the form \eqref{moments}, and moments of the form $\langle A_x^\varsigma\otimes B\rangle$, with $B$ an arbitrary operator in Bob's trusted operator algebra \cite{strings,constraint}.
All the other elements are not directly available, since they involve products of Alice's unknown operators~\cite{nonobs}, and are treated as arbitrary (complex, in general) {\it free} parameters. 

Our main goal is to check whether the observed data could be obtained or not by measurements on a separable state.
At the level of the moment matrix, assuming that the observables $A_i$ commute imposes some extra linear constraints between the elements of $\Gamma$, as discussed above. Additionally, we can also impose other constraints on $\Gamma$ given the knowledge of Bob's operators. The idea of our method then relies on searching for values for the free parameters of the constrained $\Gamma$ that make it positive semidefinite. If no such values are found, then the data are incompatible with a model relying on commuting observables on Alice's side, and consequently no separable state could give rise to it.



More formally, let $\cal R$ denote a particular simultaneous assignment of values to all independent free parameters, and let ${\bf{\Gamma}}_{\cal R}$ denote the moment matrix for commuting measurement operators on Alice's side dependent on such an assignment. Then, steering is witnessed from ${\bf{\Gamma}}_{\cal R}$ if the latter cannot be made positive semidefinite for any possible assignment $\cal R$ of the free parameters, i.e.,
\begin{equation}\label{Theorem 1}
{\bf{\Gamma}}_{\cal R} \ngeq 0, \,\,\, \forall \, {\cal R} \quad \Rightarrow \quad \mbox{$\{\langle A_x^\varsigma \otimes B_y^\tau\rangle\}$ demonstrates steering}.
\end{equation}
As anticipated, Eq.~(\ref{Theorem 1}) is the central result of this Letter.


The proposed method for investigating steerability through moments of observables shows many advantages. First, it is valid for bipartite quantum systems of any dimension, be it discrete, continuous or even hybrid since everywhere Bob's Hilbert space was assumed arbitrary, while Alice  was allowed for an arbitrary (discrete or continuous) set of outcomes.
Second, the condition \eqref{Theorem 1} serves as an infinite hierarchy of criteria; one may start with a small set of selected operators $\{ S_i \}$, that are chosen at will, and can gradually increase this set by adding more moments to improve steering detection. In particular, the operators $\{S_i\}$ can be chosen from the set $\mathcal{S}$ of all strings (products) of operators of Alice's (unknown) observables, $A_x$ and Bob's (known) observables $B_y$. This infinite set can naturally be partitioned into subsets $\mathcal{S}^{(k)}$ containing all strings of a given length $k$. For example, with only two operators on each side, $\mathcal{S}^{(0)} = \{\openone \otimes \openone\}$, $\mathcal{S}^{(1)} = \{A_1\otimes \openone, A_2\otimes \openone, \openone\otimes B_1, \openone \otimes B_2\}$,  $\mathcal{S}^{(2)} = \{A_1A_2\otimes \openone, A_2A_1 \otimes \openone, A_1\otimes B_1, A_1\otimes B_2, A_2\otimes B_1, A_2\otimes B_2, \openone \otimes B_1B_2, \openone \otimes B_2B_1\}$, etc.
Third, checking whether there is any assignment of unknown parameters which makes a matrix positive semidefinite subject to linear constraints is an instance of a semidefinite program (SDP) which can be efficiently solved for many cases of interest. Moreover, the duality theory of SDPs allows us to extract linear inequalities which act as witnesses for steering.

{\it Examples} --- In the following we consider various families of quantum states, and show that the proposed hierarchy generalizes and includes known steering criteria as special cases.

(i) \textit{ $2 \times 2$ Werner states  ---}
Consider the class of discrete variable two-qubit Werner states \cite{werner},
$
\rho_{AB}(w)= w \,|\psi^-\rangle_{AB} \langle\psi^- | + (1-w) \openone_{AB} /4 ,
$
where $|\psi^- \rangle_{AB}=\frac{1}{\sqrt{2}}\left( |01\rangle_{AB}-|10\rangle_{AB} \right)$ is the singlet. To check their steerability, we construct the moment matrix \eqref{momentmatrix} defined by  the previously mentioned set of observables ${\cal S} = \{ \openone \otimes \openone, A_1  \otimes X, A_2 \otimes Y, A_3 \otimes Z \}$ for Alice and Bob:
\begin{equation}\label{wernermatrix}
{\bf{\Gamma}}_{\cal R}  = \left( {\begin{array}{*{20}{c}}
1&{\langle {A_1 \otimes X} \rangle }&{\langle {A_2 \otimes Y} \rangle }&{\langle {A_3 \otimes Z} \rangle }\\
{\langle {A_1 \otimes X} \rangle }&{\langle {{A_1^2} \otimes {X^2}} \rangle }&{\langle {{A_1}{A_2} \otimes XY} \rangle }&{\langle {{A_1}{A_3} \otimes XZ} \rangle}\\
{\langle {A_2 \otimes Y} \rangle }&{\langle {{A_2}{A_1} \otimes YX} \rangle }&{\langle {{A_2^2} \otimes {Y^2}} \rangle }&{\langle {{A_2}{A_3} \otimes YZ} \rangle }\\
{\langle {A_3 \otimes Z} \rangle }&{\langle {{A_3}{A_1} \otimes ZX} \rangle }&{\langle {{A_3}{A_2} \otimes ZY} \rangle }&{\langle {{A_3^2} \otimes {Z^2}} \rangle }
\end{array}} \right).
\end{equation}
Consider the statistics of Alice's unknown measurements $A_1,A_2,A_3$ to originate from spin-measurements $X,Y,Z$, respectively, on her share of $\rho_{AB}$. We observe that $\langle A_1^k \otimes B \rangle = \langle X^k \otimes B \rangle_{{\rho_{AB}(w)}} $, for $k=1,2$ and arbitrary $B$, and similarly for the observable elements that contain $A_2^k$ and $A_3^k$.
Furthermore, the commutativity requirement on Alice's side, together with the algebra of operators on Bob's side (e.g. $\langle {A_1}{A_2}\otimes XY\rangle=-\langle {A_2}{A_1}\otimes YX\rangle$), reduces the number of independent free parameters to three.
One can then numerically check the positivity of the moment matrix and find that ${\bf{\Gamma}}_{\cal R} \ngeq 0, \, \forall {\cal R}$, for all $w > w_{\min}=1/\sqrt{3}$, which is known to be the threshold value for steering when Alice has exactly three inputs \cite{cavalcanti}, as is the case here. The dual of the SDP gives the following \textit{optimal steering witness}, for this family of states and measurements,
\begin{equation}\label{optwerner}
\langle A_1\otimes X\rangle +\langle A_2\otimes Y \rangle +\langle A_3 \otimes Z \rangle \geq - \sqrt{3},
\end{equation}
which is violated by all Werner states with $w > 1/\sqrt{3}$, while satisfied by all unsteerable states \cite{appendix}.
The steering criterion \eqref{optwerner} was derived independently elsewhere \cite{cavalcanti}, and we have shown that it is only a special case of our general approach.

Non-linear criteria can also be derived and, remarkably, in an analytical manner. A hermitian matrix is known to be positive semidefinite \textit{iff} all its principal minors are non-negative \cite{meyer}. Since ${\bf{\Gamma}}_{\cal R}$ is by definition hermitian, ${\bf{\Gamma}}_{\cal R} \geq 0$ implies $\det {\bf{\Gamma}}_{\cal R} \geq 0$, that can be shown to be satisfied by all unsteerable assemblages \textit{iff} \cite{appendix}
\begin{equation}\label{nonlinear}
\langle A_1\otimes X\rangle^2 +\langle A_2\otimes Y \rangle^2 +\langle A_3 \otimes Z \rangle^2 \leq 1.
\end{equation}
When applied to $\rho_{AB}(w)$, steering detection is achieved for $w$ down to the known threshold value $w_{\min}=1/\sqrt{3}$. Moreover, based on the positivity of the principal minors of \eqref{wernermatrix}, other non-linear criteria can be derived with two (instead of three) dichotomic measurements per site \cite{appendix}.

(ii) \textit{Gaussian states  ---}
Let us first make a small introduction to Gaussian states \cite{ournewreview,weedbrook}. A Gaussian state $\rho_{AB}^G$ is well-known to be fully determined up to local displacements by its covariance matrix, defined as $(\gamma_{AB})_{ij}=\mbox{Tr}[(R_i R_j+R_j R_i)\rho_{AB}^G]$, which contains all second moments for two modes. The vector $R=(q_A,p_A,q_B,p_B)^T$ conveniently groups the quadrature operators $q_{A(B)},p_{A(B)}$ for each mode, satisfying the canonical commutation relations $[R_j,R_k]=i(\Omega_{AB})_{jk}$, where $\Omega_{AB}=\Omega_A \oplus \Omega_B$ is the symplectic matrix, with $$\Omega_A=\Omega_B=
{\scriptscriptstyle \left( {\begin{array}{*{10}{c}}
0&1\\
{ - 1}&0
\end{array}} \right)}.$$ To simplify things, using local unitary operations and classical communication, we can always bring the covariance matrix into the standard form,  $$\tilde{\gamma}_{AB}=
{\scriptscriptstyle \left( {\begin{array}{*{10}{c}}
\tilde{A}&\tilde{C}\\
{ \tilde{C}^T}&\tilde{B}
\end{array}} \right)},$$ where $\tilde{A}=\mbox{diag}(a,a)$ and $\tilde{B}=\mbox{diag}(b,b)$ are the marginal covariance matrices of Alice and Bob and $\tilde{C}=\mbox{diag}(c_1,c_2)$ contains their correlations.

We proceed by investigating the steerability of Gaussian states in standard form (which implies the steerability of any non-Gaussian state with the same second moments thereof) using the following set of quadrature observables, ${\cal S} = \{A_1 \otimes \openone, A_2 \otimes \openone, \openone \otimes q_B, \openone \otimes p_B \}$, while we consider Alice's unknown measurements $A_1, A_2$ to originate from measurement of the quadratures $q_A, p_A$ respectively. The corresponding moment matrix ${\bf{\Gamma}}$ \eqref{momentmatrix} for Gaussian states in standard form becomes,
\begin{equation}\label{Gaussianmatrix}
{\bf{\Gamma}}_{\cal R} = \frac{1}{16}\left( {\begin{array}{*{20}{c}}
a&{\cal R}&{{c_1}}&0\\
{\cal R}&a&0&{{c_2}}\\
{{c_1}}&0&b&i\\
0&{{c_2}}&-i&b
\end{array}} \right),
\end{equation}
with ${\cal R} = \langle A_1 A_2 \rangle$ being the only unobservable free (real) parameter with commutativity imposed. We can proceed analytically, by remarking that if $\rho_{AB}^G$ were nonsteerable then there would exist $\cal R$ such that ${\bf{\Gamma}}_{\cal R} \geq 0$ which implies $\det {\bf{\Gamma}}_{\cal R} \geq 0$. The latter, is equivalent to $\det \tilde{\gamma}_{AB} - \det \tilde{A} \geq {\cal R}^2 (\det \tilde{B} - 1 )\geq 0$, where for the second inequality we used the property  $\det \tilde{B} \geq 1$ that all physical states must satisfy \cite{ournewreview}. Therefore, all unsteerable  assemblages necessarily satisfy $\det \tilde{\gamma}_{AB} - \det \tilde{A}\geq 0$, while a violation would signal steering since there exist no ${\cal R}$ able to make $\det {\bf{\Gamma}}_{\cal R}$ non-negative and consequently ${\bf{\Gamma}}_{\cal R}$ positive semidefinite. The steering condition $\det \tilde{\gamma}_{AB} - \det \tilde{A} \geq 0$ derived here can be shown to be satisfied \textit{iff} \cite{wiseman,Kogias},
\begin{equation} \label{Wiseman}
\tilde{\gamma}_{AB} + i (0_A \oplus \Omega_B) \geq 0,
\end{equation}
which is precisely Wiseman \textit{et al.}'s necessary and sufficient criterion for the steerability of Gaussian states under Alice's Gaussian measurements \cite{wiseman,wisepra}. Therefore, yet another  criterion turns out to be a special case of our approach and this time in the CV regime. It is worth remarking that the derivation of \eqref{Wiseman} presented here made no assumptions about either Alice's uncharacterized system or the Gaussianity of Bob's subsystem (also, see \citep{kogiasjosa} ), in contrast to \cite{wiseman}.

(iii) \textit{Lossy N00N states ---}
Consider now the following class of lossy non-Gaussian CV bipartite quantum states,
\begin{equation}\label{N00N}
\rho_{AB}^{(N)} = (1-\eta)\,\, |00\rangle_{AB} \langle 00| + \eta\,\,|N00N \rangle_{AB}\langle N00N|,
\end{equation}
where $|N00N \rangle_{AB}=\frac{1}{\sqrt{2}}\left( |N0\rangle_{AB}-|0N\rangle_{AB} \right)$ is the well-known \textit{N00N} state useful in quantum metrology \cite{NOON}, whose imperfect preparation is modelled through a mixing with the vacuum with probability  $\eta$. For later use, let us define position and momentum observables for each party $A(B)$, given $N$, as \cite{Hillery} $q_{A(B)}^{(N)}=\frac{1}{\sqrt{2}}( a_{A(B)}^{\dagger \,N} + a^N_{A(B)})$ and  $p_{A(B)}^{(N)}=\frac{i}{\sqrt{2}} \left( a_{A(B)}^{\dagger \,N} - a^N_{A(B)}\right)$, satisfying  $[ q_{A(B)}^{(N)},p_{A(B)}^{(N)}]=i$, with  $[a_{A(B)},a_{A(B)}^\dagger ]=1$.

For $N=1$, Eq.~\eqref{N00N} describes an entangled state produced by splitting a single photon (generated with probability $\eta$) at a 50-50 beam splitter.
This state is of theoretical \cite{Tan,Hardy} and experimental interest \cite{Fuwa,Lombardi}, and it is very desirable to have an experimentally friendly criterion that allows one to certify some form of nonlocality in its correlations. To our knowledge, the current best steering detection for $\rho_{AB}^{(1)}$ using only quadrature measurements is achieved by a non-linear steering inequality proposed by Jones and Wiseman \cite{JonesWise}, which can detect steering down to $\eta \geq 0.77$ in the limit of Alice having an infinite number of inputs, while both Alice and Bob bin their outcomes (i.e. for a given outcome $a$, a value is assigned 0 if $a<0$, and 1 if $a \geq 0$). For comparison, recently proposed entropic steering criteria \cite{walborn11}, employing (unbinned) quadrature measurements for both parties, can be seen to detect steering for a weaker $\eta \geq 0.94$, while all criteria that involve moments of quadratures up to second order fail to detect any steering at all \cite{reid,wiseman,cavalcanti}. We will show that our moment matrix approach outperforms all the previous methods for these states.

To make the comparison fair,  we also consider that Alice only performs two quadrature measurements, but allow Bob to measure arbitrary \textit{local} operators, see Appendix for discussion. To test for steering we use $\mathcal{S} = \{\openone\otimes \openone, A_0\otimes q_B, A_0\otimes p_B, A_1\otimes q_B, A_1\otimes p_B, A_0^2\otimes \openone, A_1^2\otimes \openone, \openone\otimes q_B^2, \openone\otimes q_Bp_B, \openone\otimes p_Bq_B, \openone\otimes p_B^2\} $, with the observable data calculated assuming Alice's unknown measurements $A_1,A_2$ are the quadratures $q_A,p_A$ respectively. Here, $q_{A(B)}$,$p_{A(B)}$ correspond to $q_{A(B)}^{(N)}$,$p_{A(B)}^{(N)}$ defined above, with $N=1$. The set $\cal S$  defines an $11 \times 11$ moment matrix ${\bf{\Gamma}}$ \eqref{momentmatrix}, with two inputs $A_1,A_2$ associated to Alice. Following the steps of the detection method,  with the observable elements of ${\bf{\Gamma}}$ computed from the state $\rho_{AB}^{(1)}$ \cite{constraint}, we employ SDP to efficiently check \eqref{Theorem 1}, and manage to detect steering for all $\eta$ down to the critical value $\eta \geq \frac{2}{3}\equiv \eta_c$, which is lower than what previous methods can achieve.
The dual of the SDP gives us the optimal linear steering inequality for $\rho_{AB}^{(1)}$, reported in the Appendix, that is violated  for all $\eta \geq \frac{2}{3}$ and satisfied by all unsteerable assemblages.
The proposed witness involves for Bob local moments of quadratures up to fourth order and can be efficiently measured by  homodyne detection and linear optics \cite{SVscheme,Avenhaus}, therefore demonstrating the experimental feasibility of our proposal.


For any given $N>1$, we can consider the same set $\cal S$, with corresponding observables $q_{A(B)}^{(N)}$, $p_{A(B)}^{(N)}$. We have tested our method up to $N=6$ and observed a steering detection down to $\eta \geq \eta_c^{(N)}$, with $\eta_c^{(N)} \lesssim \frac23$ (e.g., $\eta_c^{(6)}\approx 0.61$ for $N=6$). We conjecture that steering be detectable with our method for all $N$, although larger values could not be tested due to computational limitations. We should note however that for $N>1$ the observables $q_{A(B)}^{(N)}$, $p_{A(B)}^{(N)}$  correspond to non-Gaussian measurements that are hard to implement experimentally. On the other hand, for $N>1$ the feasible quadrature measurements $q_{A(B)}$ and $p_{A(B)}$ could not detect steering in the states of Eq.~\eqref{N00N} for any $\eta$ and for the given set $\cal S$ considered above.

\textit{Conclusion ---} We proposed an infinite hierarchy of sufficient conditions for bipartite steering applicable to all quantum systems. Other previously known steering criteria were shown to be special cases of our approach, both in the discrete and continuous variable regimes. An optimal witness for an inperfect single-photon entangled state was obtained, which was shown to be more resistant to losses than previous proposals, and experimentally accessible with linear optics and homodyne detection. In the light of a recently proved equivalence between steering and joint measurability \cite{uola,quintinojoint,uola2015one}, the hierarchy proposed here can also be used to test whether a set of Alice's inputs is not jointly measurable. An interesting future direction would be to extend the present method to multipartite steering detection, in a quantum network scenario with some trusted and some untrusted  parties \cite{cavalcanti2014detecting,he,armstrong}.

\textit{Acknowledgments ---} We acknowledge fruitful discussions with W. Vogel, M. Piani, P. Wittek, and especially J. Sperling. We thank the University of Nottingham (International Collaboration Fund), the Foundational Questions Institute (Grant No.~FQXi-RFP3-1317), the European Research Council (ERC StG GQCOP, Grant No.~637352, ERC CoG QITBOX, Grant No.~617337, ERC AdG NLST), the  Beatriu de Pin\'os fellowship (BP-DGR 2013), the EU Project SIQS, the Spanish Project FOQUS, the Generalitat de Catalunya (SGR875) and the John Templeton Foundation for financial support.


\begin{thebibliography}{62}
\bibitem [{str()}]{strings}%
  \BibitemOpen
  \href@noop {} {}\bibinfo {note} {There elements may appear non-observable,
  due to $B$ in general being non-hermitian. However, one can always find
  observables $W$ and $\Omega$ s.t., $B = W - i \Omega$, for any operator $B$.
  Henceforth, the desired element can be written as a linear combination of
  mean values of directly observable quantities.}\BibitemShut {Stop}%
\bibitem [{con()}]{constraint}%
  \BibitemOpen
  \href@noop {} {}\bibinfo {note} {From an experimental point of view, these
  values (i.e. the data) will be equal to the values observed in the
  experiment. From a theoretical point of view, when one wants to check the
  steerability of a quantum state, as we do here, these values will be equal to
  the values predicted by the state under consideration.}\BibitemShut {Stop}%
\bibitem [{non()}]{nonobs}%
  \BibitemOpen
  \href@noop {} {}\bibinfo {note} {Elements of the form $\langle A_1A_2\cdots
  \otimes B_1\rangle$ for example, are unobservable since they would require a
  simultaneous measurement of multiple inputs for Alice in order to determine
  the outcome of the product $A_1A_2\cdots$. In a steering test such
  measurements cannot be performed.}\BibitemShut {Stop}%
\end{thebibliography}

%

\clearpage

\setcounter{page}{1}
\setcounter{equation}{0}
\appendix*
\section{Appendix: Supplemental material}

\subsection{Proof of the equivalence between unsteerability and the existence of a separable model}

In the proof that follows we assume Bob's Hilbert space to be arbitrary (continuous or discrete variable), while for simplicity we assume discrete outcomes for Alice. The generalization of the proof to continuous outcomes will be immediate as we shall see.

First, we recall that Bob's assemblage $\left\lbrace \sigma^B_{a|x}\right\rbrace$, is unsteerable by Alice's inputs $x=1,...,n$ (with corresponding outcomes $a_x=1,...,d_x$) \textit{iff} it can be expressed as,
\begin{equation}\label{unsteer}
\sigma^B_{a|x}= \sum_\lambda q_\lambda p(a|x,\lambda) \, \rho_\lambda, \,\,\,\, \forall x,a.
\end{equation}
The first part of the proof amounts to expressing \eqref{unsteer} in a suitable form in terms of deterministic functions (i.e. the Kronecker delta function) that will prove very helpful.
The basic tool we utilize is the following identity,
\begin{equation}\label{determ1}
p(a|x,\lambda) = \sum_{a_x} \delta_{a,a_x}\, p(a_x|x,\lambda),
\end{equation}
for a particular input $x$, while $\delta_{i,j}$ is the Kronecker delta. By inserting in \eqref{determ1} the identities, $\sum_{a_i} p(a_i|i,\lambda)=1$, for every input $i \neq x$, we get,
\begin{equation}\label{determ2}
p(a|x,\lambda) = \sum_{a_1...a_n} \delta_{a,a_x}\, p(a_1|1,\lambda)\cdots p(a_n|n,\lambda),
\end{equation}
where the summation over $a_x$ is implicitly included.
Substituting \eqref{determ2} back in the assemblage \eqref{unsteer} we get the desired expression,
\begin{equation}\label{unsteer2}
\sigma^B_{a|x} = \sum_{a_1...a_n} \delta_{a,a_{x}} \omega_{a_1...a_n},
\end{equation}
where the unnormalized positive semidefinite operators $\omega_{a_1...a_n} \geq 0$ correspond to,
\begin{equation}\label{omega}
\omega_{a_1...a_n}=\sum_{\lambda} q_\lambda p(a_1|1,\lambda)\cdots p(a_n|n,\lambda).
\end{equation}


In the second part of the proof, we will show that one can always define a separable model $\bar{\rho}_{AB}$ for Alice and Bob, and appropriate measurement operators for Alice, that can reproduce an arbitrary unsteerable assemblage \eqref{unsteer2}.
Consider each input $x$ of Alice, with outcomes $a_x=1,...,d_x$, to correspond to a fictitious observable (hermitian operator) $A_x$ such that,
\begin{equation} \label{observable}
A_x |a_x \rangle_A = a_x |a_x \rangle_A.
\end{equation}
where the same outcomes $a_x =1,...,d_x$ correspond to its real eigenvalues with $|a_1 \rangle_A,...,|a_n \rangle_A$being the corresponding eigenvectors. When Alice announces to Bob a pair $\left(a,x \right)$, i.e.  measured input $x$ and got outcome $a$, it will be considered equivalent as if she measured the observable $A_x$ and got the eigenvalue $a$ as an outcome (with corresponding eigenvector $| a \rangle).
$ Note that such a correspondence $x \leftrightarrow A_x$ can always be made, since the announced outcomes $a_x$ always correspond to eigenvalues of some observable.

Next, \textit{assume} that all the defined observables $\{A_1,...,A_n\}$  mutually commute,
\begin{equation}\label{commute}
[A_x, A_{x'}]=0, \,\,\,\, \forall x\neq x',
\end{equation}
and, therefore, a joint basis exists that diagonalizes all $A_x, \,\forall x$, simultaneously. We will show that these commuting observables can reproduce the statistics of any unsteerable assemblage by acting on a suitable separable state. Let us denote the vectors of this basis as $\{ |a_1 \cdots a_n \rangle \}$, which sum to unity, $\sum_{a_1...a_n} |a_1 \cdots a_n\rangle_A \langle a_1 \cdots a_n | = 1$, and are orthonormal, i.e.,
\begin{equation}\label{orthonormal}
 \langle a'_1...a'_n| a_1...a_n \rangle = \delta_{a_1,a'_1}\cdots \delta_{a_n,a'_n}.
\end{equation}
Due to the simultaneous diagonalization of every observable, it holds, $A_x |a_1 \cdots a_n\rangle_A = a_x |a_1 \cdots a_n \rangle_A$, $\forall a_x$, $x$.

If $\rho_{AB}$ is the shared state between Alice and Bob, when Alice measures input $x\leftrightarrow A_x$ and announces output $a$, Bob's (unnormalized) state conditioned on the pair $(x,a)$, will be,
\begin{equation}\label{conditional}
\sigma^B_{a|x} = \mbox{Tr}_A [ (M_{a|x} \otimes 1_B) \rho_{AB} ],
\end{equation}
where we defined the projectors onto the eigenstates of $A_x$ with eigenvalue $a$,
\begin{equation}\label{projectors}
M_{a|x} =  \sum_{a_1...a_n} \delta_{a,a_x}  |a_1 \cdots a_n \rangle_A \langle a_1 \cdots a_n |,
\end{equation}
satisfying, $M_{a|x}^2=M_{a|x}$ and $\sum_{a} M_{a|x} = 1$, $\forall$ $x$. Notice the summation over the outcomes of the unannounced inputs, which is due to the inaccessibility of these degrees of freedom to Bob. Using the spectral decomposition of each $A_x$, we also get an expression for the observables, i.e.,
\begin{equation}\label{spectral}
A_x = \sum\limits_{a=1}^{d_x} a \,\, M_{a|x}.
\end{equation}

Now we will show the desired result that if $\rho_{AB}$ is the following separable state,
\begin{equation}\label{sep}
\bar{\rho}_{\rA\rB} = \sum_{a_1,\ldots,a_n} \left| a_1,\ldots,a_n\rangle_\rA\langle a_1,\ldots,a_n \right| \otimes \omega_{a_1\ldots a_n},
\end{equation}
Bob's conditional state \eqref{conditional} will correspond to the unsteerable assemblage \eqref{unsteer2} if Alice measures the commuting observables defined in \eqref{spectral}. We have,
\begin{equation}
\begin{split}
\bar{\sigma}^B_{a|x} &=  \mbox{Tr}_A [ (M_{a|x} \otimes 1_B) \bar{\rho}_{AB} ] \\
&= \sum_{a_1...a_n}\sum_{a'_1...a'_n} \delta_{a,a'_x} |\langle a'_1...a'_n| a_1...a_n \rangle|^2 \omega_{a_1...a_n}\\
&= \sum_{a_1...a_n} \delta_{a,a_x} \omega_{a_1...a_n},
\end{split}
\end{equation}
matching exactly \eqref{unsteer2}, where  we used the orthonormality of the states \eqref{orthonormal} and the property $\delta_{i,j}^2=\delta_{i,j}$ of the Kronecker delta.

The generalization of the proof from discrete to continuous outcomes for Alice is straightforward, by replacing all summations with integrals, $\sum\limits_{a_x} \rightarrow \int\limits_{-\infty}^{\infty} da_x$ and the Kronecker delta  with the Dirac delta function, $\delta_{a,a_x} \rightarrow \delta (a - a_x ),$ which is a common practice when dealing with continuous Hilbert spaces.

\subsection{SDP, Dual and Optimal steering witnesses}
First, we will show that the problem \eqref{Theorem 1} can be expressed as an SDP \cite{SDP, boydbook}, and then derive its corresponding dual problem that will lead us to the optimal steering witnesses.

Consider a square $N \times N$ (moment) matrix ${\bf{\Gamma}}$ with arbitrary elements ${\Gamma}_{ij}$.
Whether such a matrix is positive semidefinite, i.e. ${\bf{\Gamma}} \geq 0$, is equivalent to whether its smallest eigenvalue, $\lambda_\star$, is non-negative, i.e. $\lambda_\star \geq 0$.
The steering detection method outlined in the main Letter, boils down to finding the maximized $\lambda_\star$ (name it, $\lambda^{\max}_\star$) over all possible (complex, in general) values of the moment matrix's elements $\{{\Gamma}_{ij}\}$ satisfying at the same time two types of constraints:
\\
(a) All the observable elements of ${\bf{\Gamma}}$ are constrained to be equal to the observable values from the steering test \cite{constraint}.
\\
(b) Linear relations between the unobservable elements, imposed by the commutativity constraint of Alice's operators and the utilization of Bob's operator algebra.

The semidefinite program corresponding to the problem described takes the following standard form \cite{SDP},
\begin{equation} \label{SDPgamma}
\begin{split}
\lambda_{\star}^{\text{max}} \,\,\,= \,\,\,& \max\limits_{\lambda, \{\Gamma_{ij}\}} \,\,\,\,\,\,\,\,\,\,\,\,\,\,\,\, \lambda \\
&\mbox{subject to} \,\,\,\,\,\,\, {\bf{\Gamma}}-\lambda\, 1\geq 0 \\
&\,\,\,\,\,\,\,\,\,\,\,\,\,\,\,\,\,\,\,\,\,\,\,\,\,\,\,\,\,\mbox{Tr}\left[{\bf{\Gamma}} A_i \right] = b_i, \,\,\, i=1,...,k \\
&\,\,\,\,\,\,\,\,\,\,\,\,\,\,\,\,\,\,\,\,\,\,\,\,\,\,\,\,\,\mbox{Tr}\left[{\bf{\Gamma}} C_j \right] = 0, \,\,\,\, j=1,...,l
\end{split}
\end{equation}
known as the \textit{primal problem}, the output of which will be $\lambda_{\star}^{\text{max}}$.
The first constraint in \eqref{SDPgamma} guarantees that the output of the SDP will be equal to the smallest eigenvalue of the given ${\bf{\Gamma}}$.
The second and third constraints correspond to the constraints (a) and (b) respectively, with suitably chosen matrices $A_i$ and $C_j$ depending on the particular ${\bf{\Gamma}}$, while $k$ and $l$ correspond to the total number of observable elements and linear relations respectively. The values $b_i$ are the ones obtained from the steering test, as explained in \cite{constraint}.
Concluding, steering will be witnessed from the SDP \eqref{SDPgamma} if $\lambda_{\star}^{\max} < 0$.

To obtain the dual of the SDP \eqref{SDPgamma}, the solution of which will give us an upper bound on the quantity of interest $\lambda_\star^{\max}$, we start by writing the Lagrangian of this problem \cite{boydbook},
\begin{equation}\label{lagrangian}
\begin{split}
{\cal L} &= \lambda + \Tr [\, Z \cdot ( {\bf{\Gamma}} - \lambda\, 1 ) \, ] + \\
&\,\,\,\,\,\,\,\,\,\,\,\,\,\,\,+ \sum\limits_{i=1}^{k} \mu_i^* \left( b_i - \Tr[{\bf{\Gamma}} \cdot A_i ] \right) + \sum\limits_{j=1}^{l} \nu_j^* \left(0 - \Tr [ {\bf{\Gamma}} \cdot C_j ]\right)\\
&=  \sum\limits_{i=1}^k \mu_i^* b_i + \lambda\left(1-\Tr Z\right) + \Tr \left[ {\bf{\Gamma}} \cdot \left( Z -\sum\limits_{i=1}^k \mu_i^* A_i - \sum\limits_{j=1}^l \nu_j^* C_j  \right) \right]
\end{split}
\end{equation}
where the $N \times N$ hermitian matrix $Z$ and the complex variables $\{\mu_i\}$ and $\{\nu_j\}$ are the \textit{dual variables} to the first, second and third (sets of) constraints in \eqref{SDPgamma} respectively.  If we consider the maximized value $\max\limits_{\lambda,\{{\bf{\Gamma}}\}_{ij}} {\cal L}$ over the primal variables $\lambda,\{{\bf{\Gamma}}\}_{ij}$, it's straightforward to see from \eqref{lagrangian} that, $\max\limits_{\lambda,\{{\bf{\Gamma}}\}_{ij}} {\cal L} \geq  \lambda^{\max}_\star +  \Tr [\, Z \cdot ( {\bf{\Gamma}} -  \lambda^{\max}_\star\, \openone ) \, ] $.
Therefore choosing $Z \geq 0$, and since ${\bf{\Gamma}} -  \lambda^{\max}_\star\ \openone \geq 0$ due to the first constraint in \eqref{SDPgamma}, we find the following bound,
\begin{equation}\label{bound}
\max\limits_{\lambda,\{{\Gamma}_{ij}\}} {\cal L} \geq \lambda^{\max}_\star ,
\end{equation}
Our goal is to use $\max\limits_{\lambda,\{{\Gamma}_{ij}\}} {\cal L}$ to get a good estimate for the figure of merit $\lambda^{\max}_\star $, and in order for the bound \eqref{bound} not to be trivial (i.e. equal to infinity), $\cal L$ should be bounded from above. We see that this occurs trivially if we set as constraints for the dual variables,
\begin{gather}
\Tr\, Z = 1 \\
Z = \sum\limits_{i=1}^k \mu_i^* A_i + \sum\limits_{j=1}^l \nu_j^* C_j,
\end{gather}
in addition to $Z \geq 0$. Imposing these constraints on $\cal L$, the Lagrangian \eqref{lagrangian} optimized over the primal variables takes the simple form,
\begin{equation}\label{bound2}
\max\limits_{\lambda,\{{\Gamma}_{ij}\}} {\cal L} = \sum\limits_{i=1}^k \mu_i^* b_i \geq  \lambda^{\max}_\star.
\end{equation}
Therefore we are lead to an alternative approach to bound the desired quantity $\lambda^{\max}_\star$, by minimizing the left-hand side over the dual variables, for given $\{b_i\}$, leading us to the following \textit{dual problem},
\begin{equation} \label{dual}
\begin{split}
\beta_\star \,\,\,\,\,\,\,= &\min\limits_{\{\mu_i\},\{\nu_j\},\{Z_{ij}\}} \,\,\,\,\, \sum\limits_{i=1}^k \mu_i^* b_i \\
&\mbox{subject to} \,\,\,\,\,\,\, Z \geq 0 \\
&\,\,\,\,\,\,\,\,\,\,\,\,\,\,\,\,\,\,\,\,\,\,\,\,\,\,\,\,\,\mbox{Tr}Z = 1, \\
&\,\,\,\,\,\,\,\,\,\,\,\,\,\,\,\,\,\,\,\,\,\,\,\,\,\,\,\,\,Z = \sum\limits_{i=1}^k \mu_i^* A_i + \sum\limits_{j=1}^l \nu_j^* C_j.
\end{split}
\end{equation}

The output of the dual \eqref{dual}, $\beta_\star$, is the tightest upper bound to the figure of merit $\lambda^{\max}_\star$  \eqref{bound2} since optimal coefficients $\{ {\bar \mu}_i \}$ are found for the given observable values $\{ b_i \}$. A negative value, $\beta_\star < 0$, is a sufficient condition for steerability, since it would imply that  $\lambda^{\max}_\star <0$, while a non-negative value $\beta_\star \geq 0$ is obtained for all unsteerable assemblages. Also, note that mere knowledge of the dual matrix $Z$ (output of \eqref{dual}) and the moment matrix $\bf \Gamma$ is enough to find $\beta_\star$ since, $\Tr \left[ {\bf \Gamma} Z
\right] = b_\star$, due to the second and third constraints in \eqref{SDPgamma}. To generalize this witness to any system, and therefore to arbitrary observations, consider arbitrary observable values $\{\bar b_i \}\neq \{ b_i \}$ but keep the same coefficients $\{ {\bar \mu}_i \}$ as before. The following linear inequality, or \textit{steering witness},
\begin{equation}\label{witnessUS}
\sum\limits_{i=1}^k {\bar \mu}_i^* \bar b_i \geq 0,
\end{equation}
is satisfied by all unsteerable assemblages while a violation signals steering detection. For the particular $\{ b_i \}$  the violation of \eqref{witnessUS} is maximal since the coefficients $\{{\bar \mu}_i \}$ are optimal for these particular values and non-optimal for any other, and therefore we refer to \eqref{witnessUS} as the \textit{optimal steering witness} for the values $\{\bar b_i \} = \{ b_i \}$, obtained by particular measurements and assemblages.

Finally, it is easy to verify that the primal problem is strictly feasible -- i.e. there exists a ${\bf{\Gamma}}$ satisfying all the equality constraints which is strictly positive definite. As such, strong duality holds for the primal and dual SDP problems, such that the optimal value of the primal $\lambda^{\max}_\star$ and the optimal value of the dual $\beta_\star$ are equal.

\subsection{Analytical derivation of non-linear steering criteria}

Consider the moment matrix ${\bf{\Gamma}}_{\cal R}$ \eqref{wernermatrix} obtained by the set of measurements, ${\cal S} = \{ \openone \otimes \openone, A_1  \otimes X, A_2 \otimes Y, A_3 \otimes Z \}$, where the statistics of Alice's unknown measurements $A_1,A_2,A_3$ also originate from ``spin''-measurements $X,Y,Z$.
In the following derivation, only the algebra of Alice's and Bob's observables will matter, independently of their shared state $\rho_{AB}$. Applying the steps of the \textit{detection method}, i.e. commutativity and the operator algebra on Bob's side, the matrix \eqref{wernermatrix} can be seen to get the  simple form,
\begin{equation}\label{nlinearmatrix}
{\bf{\Gamma}}_{\cal R}  = \left( {\begin{array}{*{20}{c}}
1&{\left\langle {A_1 \otimes X} \right\rangle }&{\left\langle {A_2 \otimes Y} \right\rangle }&{\left\langle {A_3 \otimes Z} \right\rangle }\\
{\left\langle {A_1 \otimes X} \right\rangle }&{1 }&{i\, R_1 }&{i\, R_2}\\
{\left\langle {A_2 \otimes Y} \right\rangle }&{-i\,R_1}&{1 }&{i\,R_3}\\
{\left\langle {A_3 \otimes Z} \right\rangle }&{-i\, R_2}&{-i\,R_3}&{1}
\end{array}} \right),
\end{equation}
where the three free parameters $R_i$ are real, and equal to,
$
R_1 = \langle A_1 A_2 \otimes Z \rangle,\,
R_2 = \langle A_2 A_3 \otimes X \rangle,$ and $
R_3 = - \langle A_1 A_3 \otimes Y \rangle$.
Notice that the diagonal observable terms are equal to unity independently of the shared state, due to the fact that the Pauli operators, and the observables of Alice, take values $\pm 1$, and therefore square to the identity.

As explained in the main text, the necessary condition for unsteerability ${\bf{\Gamma}}_{\cal R} \geq 0$ implies the following conditions for its principal minors,
\begin{equation}\label{minorG}
\begin{split}
\det {\bf{\Gamma}}_{\cal R} = 1- & \left\langle {A_1 \otimes X} \right\rangle^2 - \left\langle {A_2 \otimes Y} \right\rangle^2 - \left\langle {A_3 \otimes Z} \right\rangle^2 \\
&+ f\left( R_1,R_2,R_3 \right) \geq 0,
\end{split}
\end{equation}
\begin{gather}\label{minorP2}
\det P_2 = 1 - \left\langle {A_2 \otimes Y} \right\rangle^2 - \left\langle {A_3 \otimes Z} \right\rangle^2 - R_3^2 \geq 0,\\
\label{minorP3}
\det P_3 = 1 - \left\langle {A_1 \otimes X} \right\rangle^2 - \left\langle {A_3 \otimes Z} \right\rangle^2 - R_2^2 \geq 0,\\
\label{minorP4}
\det P_4 = 1 - \left\langle {A_1 \otimes X} \right\rangle^2 - \left\langle {A_2 \otimes Y} \right\rangle^2 - R_1^2 \geq 0,
\end{gather}
with,
\begin{equation}\label{functionf}
\begin{split}
 f\left( R_1,R_2,R_3 \right) = & \left( R_3 \left\langle {A_1 \otimes X} \right\rangle - R_2 \left\langle {A_2 \otimes Y} \right\rangle + R_1 \left\langle {A_3 \otimes Z} \right\rangle \right)^2\\
 & - R_1^2-R_2^2-R_3^2,
\end{split}
\end{equation}
where  the matrix $P_i$ is obtained by ${\bf{\Gamma}}_{\cal R}$ by deleting its $i$-th row and column. Each of the  conditions \eqref{minorP2}-\eqref{minorP4} leads to a steering criterion. For example,
\begin{equation}\label{detP2}
\det P_2 \geq 0\,\, \Rightarrow \,\, 1 - \left\langle {A_2 \otimes Y} \right\rangle^2 - \left\langle {A_3 \otimes Z} \right\rangle^2 \geq R_3^2 \geq 0,
\end{equation}
and similarly for \eqref{minorP3},\eqref{minorP4}. A violation of the last inequality in \eqref{detP2} signals steering since there exist no assignment for the free parameters $R_i$ that can make \eqref{detP2} non-negative.  When applied to the family of Werner states these criteria can be seen to detect steering for $w > \frac{1}{\sqrt{2}}$, which is a weaker detection than what the optimal witness \eqref{optwerner} and the stronger non-linear criterion \eqref{nonlinear} can achieve. This is of course to be expected, since the former criteria only involve two measurement settings per site.

The stronger non-linear criterion \eqref{nonlinear}, based on three measurement settings, can be derived from \eqref{minorG}, where the contribution of the free parameters is grouped in the function $ f\left( R_1,R_2,R_3 \right)$.   Our goal is to provide an upper bound for this function, say $f \leq f_{\max} $, and therefore limit its capability of making \eqref{minorG} positive for any given measurements. As a simple example of the logic behind, the analogous function in \eqref{minorP2} would be  $-R_3^2$ and is upper bounded by zero, as seen in the steering criterion \eqref{detP2}.
The maximum of $ f\left( R_1,R_2,R_3 \right)$ can be seen to  correspond to the following  values for $R_1,R_2$,
\begin{gather}\label{R1star}
\left. \partial_{R_1} f =0\right|_{R_1=R_1^\star} \,\, \Rightarrow \,\, R_1^\star = R_3 \frac{\left\langle {A_1 \otimes X} \right\rangle\left\langle {A_3 \otimes Z} \right\rangle}{\det P_2 +R_3^2}\\ \label{R2star}
\left. \partial_{R_2} f =0\right|_{R_2=R_2^\star} \,\, \Rightarrow \,\, R_2^\star = - R_3 \frac{\left\langle {A_1 \otimes X} \right\rangle\left\langle {A_2 \otimes Y} \right\rangle}{\det P_2 +R_3^2}.
\end{gather}
Therefore,
\begin{equation}\label{upperbound}
\begin{split}
f\left( R_1,R_2,R_3 \right)& \leq f\left( R_1^\star,R_2^\star,R_3 \right)\\
& = - R_3^2\, \frac{1- \left\langle {A_1 \otimes X} \right\rangle^2 - \left\langle {A_2 \otimes Y} \right\rangle^2 - \left\langle {A_3 \otimes Z} \right\rangle^2}{\det P_2 +R_3^2}.
\end{split}
\end{equation}

We employ this  bound in \eqref{minorG} and find that unsteerability of Bob's assemblage implies,
\begin{equation} \label{crit}
\begin{split}
& \det {\bf{\Gamma}}_{\cal R} \geq 0 \,\,\,  \Rightarrow \\
& 1- \left\langle {A_1 \otimes X} \right\rangle^2 - \left\langle {A_2 \otimes Y} \right\rangle^2 - \left\langle {A_3 \otimes Z} \right\rangle^2 + f\left( R_1^\star,R_2^\star,R_3 \right) \geq 0\\
& \Leftrightarrow  \left(  1- \left\langle {A_1 \otimes X} \right\rangle^2 - \left\langle {A_2 \otimes Y} \right\rangle^2 - \left\langle {A_3 \otimes Z} \right\rangle^2 \right) \frac{\det P_2}{\det P_2 +R_3^2} \geq 0
\end{split}
\end{equation}
Unsteerable assemblages necessarily satisfy $\det P_2 \geq 0$ (see \eqref{minorP2}), and therefore the last inequality of  \eqref{crit} implies the desired non-linear criterion \eqref{nonlinear},
\begin{equation}\label{nonlinear2}
\left\langle {A_1 \otimes X} \right\rangle^2 + \left\langle {A_2 \otimes Y} \right\rangle^2 + \left\langle {A_3 \otimes Z} \right\rangle^2 \leq 1.
\end{equation}
Notice that for the expressions \eqref{R1star}, \eqref{R2star} we have assumed, $\left| \langle A_2 \otimes Y \rangle \right| < 1$ and $\left| \langle A_3 \otimes Z \rangle \right| < 1$. The cases where equality is attained in either (or both) inequalities should be treated separately, and it's straightforward to see that in every single case the same condition \eqref{nonlinear2} is always obtained. Therefore, the validity of \eqref{nonlinear2} extends to the whole range of possible experimental outcomes.

\subsection{Optimal Witness for Lossy Single Photon state}
In this appendix we  provide the optimal steering witness which certifies the steerability of the noisy single photon state. As described in the main text, we used the $11 \times 11$ moment matrix defined by the set of operators $\mathcal{S} = \{\openone\otimes \openone, A_0\otimes q_B, A_0\otimes p_B, A_1\otimes q_B, A_1\otimes p_B, A_0^2\otimes \openone, A_1^2\otimes \openone, \openone\otimes q_B^2, \openone\otimes q_Bp_B, \openone\otimes p_Bq_B, \openone\otimes p_B^2\} $. First, note that moments of the form $\langle A_x^k \otimes B \rangle$ appearing in the moment matrix, with $B$ an arbitrary string of length 2 or more, are expected in general to be hard to measure experimentally. In the following we therefore assume these terms to be unobservable (and therefore treat them as free parameters in the moment matrix), and apply only the operator algebra of Bob to place linear relations between them. On the other hand, local moments of the form $\langle \openone \otimes B\rangle $ can be measured efficiently by Bob, for example by estimating his local Wigner function or by using a linear optics scheme proposed by Shchukin and Vogel \cite{SVscheme}, and therefore we keep these moments as observable. The freedom that the method gives us to keep only those measurements that can be  efficiently performed as observable, highlights the flexibility of our approach to maintain experimental feasibility. Our ultimate goal is to provide an experimentally-friendly optimal steering witness.

The code was implemented using {\sc cvx} for {\sc matlab} \cite{CVX}, with the optimal inequality extracted by solving the primal \eqref{SDPgamma} and dual \eqref{dual} problems. The optimal inequality \eqref{witnessUS} for the noisy single photon state with $\eta = 0.67$  is given by
\begin{widetext}
\begin{multline}\label{e:ineq}
\beta =  8.1657 -(\Expt{A_0}{q_B}+\Expt{A_1}{p_B})
+0.2508\,(\Expt{A_0}{q_B^3}+\Expt{A_1}{p_B^3})
-0.3110\,(\Exp{A_0^2} + \Exp{A_1^2})\\
+0.3205\,(\Expt{A_0^2}{q_B^2} + \Expt{A_1^2}{p_B^2})
+0.3020\,(\Expt{A_0^2}{p_B^2} + \Expt{A_1^2}{q_B^2})
-0.0001\,(\Expt{A_0^3}{q_B} + \Expt{A_1^3}{p_B})\\
+ 7.7217\,(\Exp{q_B^4}+\Exp{p_B^4}) + 15.5451\, \Exp{q_B^2 p_B^2} - 31.0941\,(\Exp{q_B^2}+\Exp{p_B^2}) - 31.0903 i\, \Exp{q_B p_B}
 \geq 0,
\end{multline}
\end{widetext}
satisified by all unsteerable assemblages, with the state numerically achieving the violation $\beta = -8.88\times 10^{-4}$, which is (in magnitude) far above the numerical precision. Smaller values of $\eta$ still show a violation, with numerical evidence suggesting all $\eta > 2/3$ demonstrate steering.  The maximum violation of the inequality is $\beta_{\max} = -0.1556$, achieved for $\eta = 1$.

Let us now comment on the experimental feasibility for the estimation of the witness \eqref{e:ineq}. Most of the terms in Eq. \eqref{e:ineq} can be efficiently measured  by performing homodyne detection. The term that provides some extra difficulty in its measurement is the local fourth-order moment $\langle q_B^2 p_B^2 \rangle$ of Bob. As mentioned before, for the estimation of this term Bob could implement tomography on his local state, which doesn't require conditioning on Alice's outcomes. A more efficient approach that avoids tomography would be to use a  scheme  proposed by Shchukin and Vogel \cite{SVscheme}, based on linear optics, that was designed to measure such local moments. A similar scheme was recently implemented by Avenhaus \textit{et al.}  \cite{Avenhaus}, who managed to accurately measure moments of a single-mode up to eighth order. Therefore, we can safely conclude that the proposed steering witness \eqref{e:ineq} can be efficiently measured in the laboratory.

Finally, let us note that the only terms which appear in the inequality are those which were considered observable in the moment matrix. However, observable terms of the form $\langle A_x^k \otimes B\rangle$, which are experimentally demanding, were considered unobservable, and as one would expect steering detection weakens due to such relaxation. If on the other hand we consider all these experimentally demanding terms to be observable, we find the same critical noise $\eta > 2/3$, with only the magnitude of the violation increasing (and the inequality containing the additional observable terms absent in \eqref{e:ineq}).

\end{document}